\documentclass[twoside]{ilcws07}
\usepackage[latin1]{inputenc}
\usepackage[dvips]{graphicx,epsfig,color}
\usepackage{wrapfig,rotating}
\usepackage{amssymb,amsmath,array}

\input{paperdef}

\begin{document}

\thispagestyle{empty}
\setcounter{page}{0}
\def\thefootnote{\fnsymbol{footnote}}

\begin{flushright}
\mbox{}
\end{flushright}

\vspace{1cm}

\begin{center}

{\large\sc {\bf Interplay of Electroweak Precision Observables\\[.5em]
and \boldmath{$B$} Physics Observables}}
\footnote{talk given at the {\em LCWS07}, 
May 2007, DESY, Hamburg, Germany}

\vspace{1cm}

{\sc 

S.~Heinemeyer
\footnote{
email: Sven.Heinemeyer@cern.ch}

}

\vspace*{1cm}

{\it
Instituto de Fisica de Cantabria (CSIC-UC), 
Santander,  Spain 
}
\end{center}

\vspace*{0.2cm}

\BC {\bf Abstract} \EC
Indirect information about the possible scale of
supersymmetry (SUSY) breaking is provided by $B$-physics observables (BPO) as
well as electroweak precision observables (EWPO). We review the 
combination of the constraints imposed by recent measurements of the  BPO
$\br(b \to s \ga)$, $\br(B_s \to \mu^+\mu^-)$, 
$\br(B_u \to \tau \nu_\tau)$ and $\De M_{B_s}$ with those obtained
from the experimental measurements of 
the EWPO $\MW$, $\sweff$, $\Ga_Z$, $(g-2)_\mu$ and $\Mh$.
We perform a $\chi^2$~fit to the parameters of the constrained minimal
supersymmetric extension of the Standard Model (CMSSM), in which the
SUSY-breaking parameters are universal at the GUT scale.
Assuming that the lightest supersymmetric particle (LSP) provides the
cold dark matter density preferred by WMAP and other cosmological data,
we confirm the preference found previously for a
relatively low SUSY-breaking scale, though there is some slight tension
between the EWPO and the BPO.

\def\thefootnote{\arabic{footnote}}
\setcounter{footnote}{0}

\newpage


\graphicspath{{figs/}}

\title{
Interplay of Electroweak Precision Observables\\ 
and \boldmath{$B$} Physics Observables} 
\author{S.~Heinemeyer
\vspace{.3cm}\\
Instituto de Fisica de Cantabria (CSIC-UC), Santander,  Spain 
}

\maketitle

\begin{abstract}
Indirect information about the possible scale of
supersymmetry (SUSY) breaking is provided by $B$-physics observables (BPO) as
well as electroweak precision observables (EWPO). We review the 
combination of the constraints imposed by recent measurements of the  BPO
$\br(b \to s \ga)$, $\br(B_s \to \mu^+\mu^-)$, 
$\br(B_u \to \tau \nu_\tau)$ and $\De M_{B_s}$ with those obtained
from the experimental measurements of 
the EWPO $\MW$, $\sweff$, $\Ga_Z$, $(g-2)_\mu$ and $\Mh$.
We perform a $\chi^2$~fit to the parameters of the constrained minimal
supersymmetric extension of the Standard Model (CMSSM), in which the
SUSY-breaking parameters are universal at the GUT scale.
Assuming that the lightest supersymmetric particle (LSP) provides the
cold dark matter density preferred by WMAP and other cosmological data,
we confirm the preference found previously for a
relatively low SUSY-breaking scale, though there is some slight tension
between the EWPO and the BPO. 

\end{abstract}

\section{Introduction}

In order to achieve a simplification of the plethora of soft
SUSY-breaking parameters appearing in the general MSSM, 
one assumption that is frequently employed is
that (at least some of) the soft SUSY-breaking parameters are universal
at some high input scale, before renormalization. 
One model based on this simplification is the 
constrained MSSM (CMSSM), in which all the soft SUSY-breaking scalar
masses $m_0$ are assumed to be universal at the GUT scale, as are the
soft SUSY-breaking gaugino masses $m_{1/2}$ and trilinear couplings
$A_0$. Further parameters are $\tb$, the ratio of the two vacuum
expectaion values, and the sign of the Higgs mixing parameter $\mu$. 

Within the CMSSM we perform a combined $\chi^2$~analysis~\cite{ehoww} of
electroweak precision observables (EWPO)~\cite{PomssmRep}, going
beyond previous such  
analyses~\cite{ehow3,ehow4} 
(see also \citere{other}), and of $B$-physics observables (BPO),
including some that have not been
included before in comprehensive analyses of the SUSY parameter
space (see, however, \citere{LSPlargeTB}). 
The set of EWPO included in the analysis is
the $W$~boson mass $\MW$,  the effective leptonic weak mixing angle
$\sweff$, the total $Z$~boson width
$\Ga_Z$, the anomalous magnetic moment of the  muon $(g-2)_\mu$, and
the mass of the lightest MSSM Higgs boson mass $\Mh$. In addition, we
include four BPO:  the branching ratios 
$\br(b \to s \ga)$, $\br(B_s \to \mu^+ \mu^-)$ 
and $\br(B_u \to \tau \nu_\tau)$, and the $B_s$ mass mixing parameter
$\De M_{B_s}$. 
For the evaluation of the BPO we assume minimal flavor violation (MFV)
at the electroweak scale. 


\section{The \boldmath{$\chi^2$} evaluation}

Assuming that the nine observables listed above are
uncorrelated, a $\chi^2$ fit 
has been performed with
\BE
\chi^2 \equiv \sum_{n=1}^{7} \KKL \KL
              \frac{R_n^{\rm exp} - R_n^{\rm theo}}{\si_n} \KR^2
              + 2 \log \KL \frac{\si_n}{\si_n^{\rm min}} \KR \KKR
                                               + \chi^2_{\Mh}
                                               + \chi^2_{B_s}.
\label{eq:chi2}
\EE
Here $R_n^{\rm exp}$ denotes the experimental central value of the
$n$th observable ($\MW$, $\sweff$, $\Ga_Z$, \mbox{$(g-2)_\mu$} and
$\br(b \to s \ga)$, $\br(B_u \to \tau \nu_\tau$), $\De M_{B_s}$),
$R_n^{\rm theo}$ is the corresponding MSSM prediction and $\si_n$
denotes the combined error (intrinsic, parametric (from $\mt$, $\mb$,
$\als$, $\De\al_{\rm had}$), and experimental).
Additionally,
$\si_n^{\rm min}$ is the minimum combined error over the parameter space of
each data set as explained below, and
$\chi^2_{\Mh}$ and $\chi^2_{B_s}$ denote the $\chi^2$ contribution
coming from the experimental limits 
on the lightest MSSM Higgs boson mass and on $\br(B_s \to \mu^+\mu^-)$,
respectively, see \citere{ehoww} for details.

In order to take the $\mt$ and $\mb$ parametric uncertainties correctly into
account, we evaluate the SUSY spectrum and the observables for
each data point first for the nominal values 
$\mt = 171.4 \gev$~\cite{mt1714}%
\footnote{Using the most recent experimental value, 
$\mt = 170.9 \pm 2.1 \gev$~\cite{mt1709} would have a minor impact on our
analysis.}%
~and $\mb(\mb) = 4.25 \gev$,
then for $\mt = (171.4 + 1.0) \gev$ and $\mb(\mb) = 4.25 \gev$,
and finally for $\mt = 171.4 \gev$ and $\mb(\mb) = (4.25 + 0.1) \gev$.
The latter two evaluations are used by appropriate rescaling to estimate
the full parametric uncertainties induced by the experimental uncertainties
$\de\mt^{\rm exp} = 2.1 \gev$~\cite{mt1714}
and $\de\mb(\mb)^{\rm exp} = 0.11 \gev$.
These parametric uncertainties are then added to the other errors
(intrinsic, parametric ($\als$, $\De\al_{\rm had}$), and experimental).

In regions that depend sensitively on the input values of $\mt$ and
$\mb(\mb)$, such as the focus-point region~\cite{focus} in the CMSSM, the
corresponding parametric uncertainty can become very large. In essence, the
`WMAP hypersurface' moves significantly as $\mt$ varies (and to a lesser extent
also $\mb(\mb)$), but remains thin. Incorporating this large
parametric uncertainty naively in \refeq{eq:chi2} would artificially
suppress the
overall $\chi^2$ value for such points. This artificial suppression is
avoided by adding the second term in \refeq{eq:chi2}, where 
$\si_n^{\rm min}$ is the value of the combined error evaluated for
parameter choices which minimize $\chi^2_n$ over the full data set.

Throughout this analysis, we focus our attention on parameter points
that yield the correct value of the cold dark matter density inferred
from WMAP and other data, namely 
$0.094 < \Omega_{\rm CDM} h^2 < 0.129$~\cite{WMAP}. 
The fact that the density is relatively well known
restricts the SUSY parameter space to a thin, fuzzy `WMAP hypersurface',
effectively reducing its dimensionality by one. The variations in the
EWPO and BPO across this hypersurface may in general be neglected, so
that we may 
treat the cold dark matter constraint effectively as a $\delta$
function. 
We note, however, that for any given value of $m_{1/2}$ there may be
more than one value of $m_0$ that yields a cold dark matter density
within the allowed range, implying that there may be more than one WMAP
line traversing the the $(m_{1/2}, m_0)$ plane.
 Specifically, in the CMSSM there is, in general, one WMAP line in the
coannihilation/rapid-annihilation funnel region and another in the
focus-point region, at higher $m_0$. Consequently, each EWPO and BPO may
have more than one value for any given value of $m_{1/2}$. In the
following, we restrict our study of the upper WMAP line to the part with
$m_0 < 2000 \gev$ for $\tb = 10$ and $m_0 < 3000 \gev$ for 
$\tb = 50$, restricting in turn the range of $m_{1/2}$. 

For our CMSSM analysis, the fact that the cold dark matter density is
known from astrophysics and cosmology with an uncertainty smaller
than~$10~\%$ fixes with proportional precision one combination of the
SUSY parameters, enabling us to analyze the 
overall $\chi^2$~value as a function of
$m_{1/2}$ for fixed values of $\tb$ and $A_0$. The value of $|\mu|$ is
fixed by the electroweak vacuum conditions (and $\mu > 0$ due to
$(g-2)_\mu$), the value of $m_0$ is fixed 
with a small error by the dark matter density, and the Higgs mass
parameters are fixed by the universality assumption. As in previous
analyses, we consider various representative values of 
$A_0 \propto m_{1/2}$  for the specific choices $\tb = 10, 50$.


\section{The \boldmath{$\chi^2$} analyses for EWPO, BPO and combined}

\begin{figure}[tbh!]
\begin{center}
\includegraphics[width=.48\textwidth,height=4.5cm]{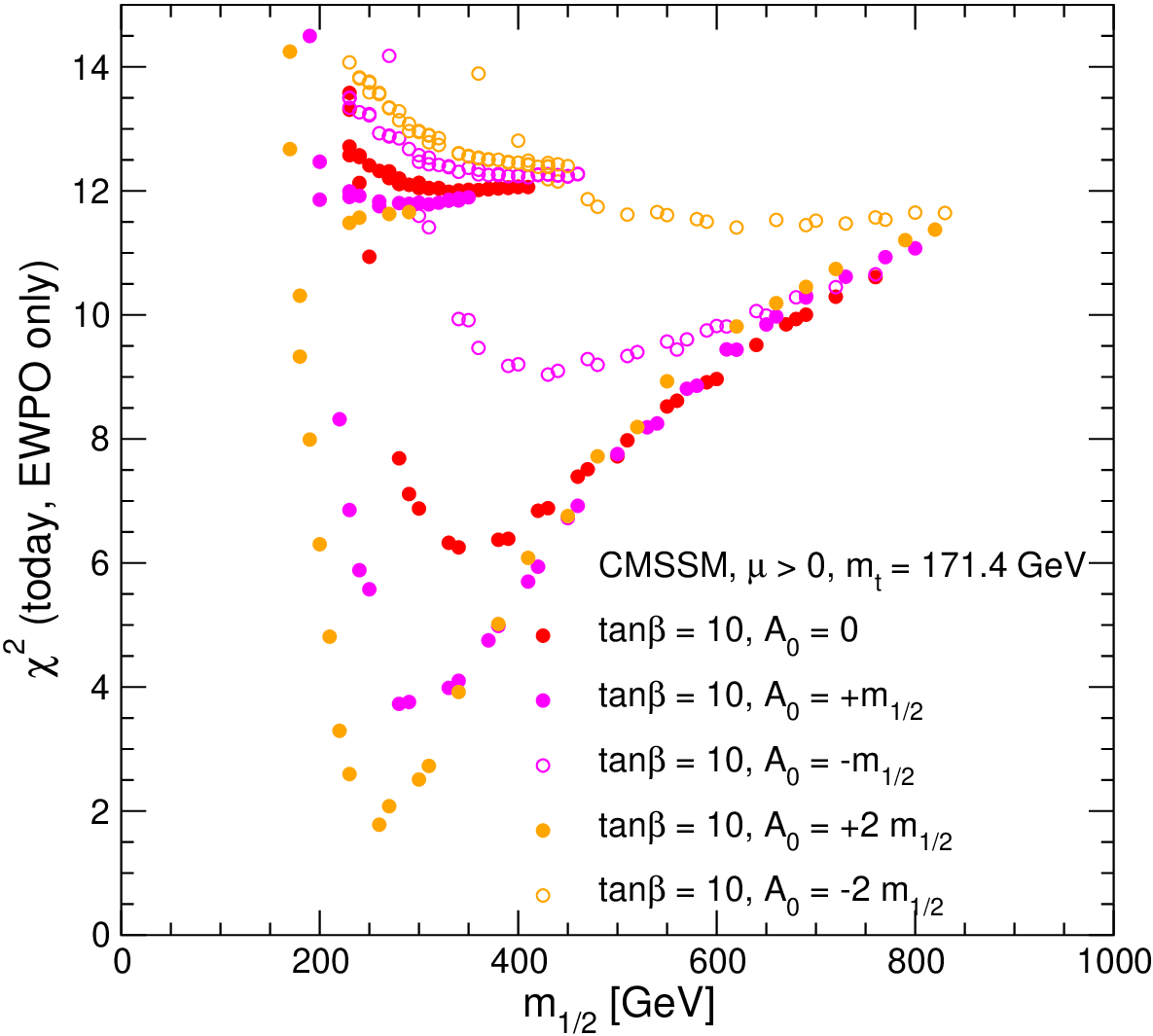}
\includegraphics[width=.48\textwidth,height=4.5cm]{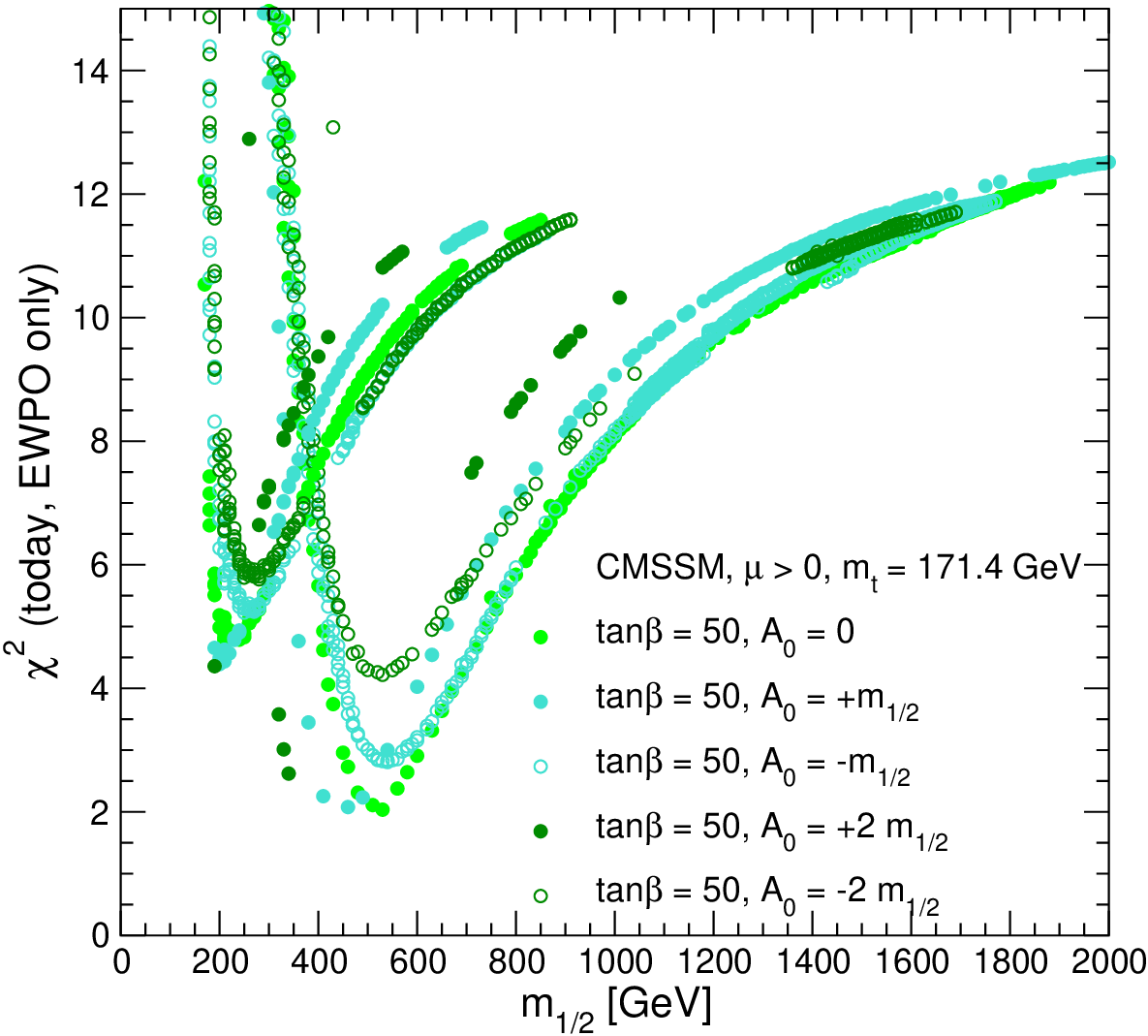}
\vspace{-0.5em}
\caption{%
The combined $\chi^2$~function for the electroweak
observables $\MW$, $\sweff$, $\Ga_Z$, $(g - 2)_\mu$ and $\Mh$, 
evaluated in the CMSSM for $\tb = 10$ (left) and
$\tb = 50$ (right) for various discrete values of $A_0$.
}
\label{fig:chi_ewpo}
\end{center}
\vspace{-0.0em}
\end{figure}

Here we show the $\chi^2$~results as 
a function of $m_{1/2}$, using \refeq{eq:chi2}. As
a first step, Fig.~\ref{fig:chi_ewpo} displays the $\chi^2$~distribution
for the EWPO alone. 
In the case $\tb = 10$ (left panel), we see a well-defined minimum of 
$\chi^2$ for $m_{1/2} \sim 300 \gev$ when $A_0 > 0$, which disappears
for large negative $A_0$ and is not present in the focus-point
region. The rise at small $m_{1/2}$ is due both to the lower limit on
$\Mh$ coming from the direct search at LEP~\cite{LEPHiggs} and to 
$(g - 2)_\mu$, whilst 
the rise at large $m_{1/2}$ is mainly due to $(g - 2)_\mu$.
The measurement of $\MW$ leads to a slightly lower minimal value of~$\chi^2$,
but there are no substantial contributions from
any of the other EWPO. The preference for $A_0 > 0$ in the
coannihilation region is due to $\Mh$, see the left plot in
\reffi{fig:Mh}, and the 
relative disfavor for the focus-point regions is due to its mismatch
with $(g - 2)_\mu$.
In the case $\tb = 50$ (right panel), we again
see a well-defined minimum of $\chi^2$, this time for $m_{1/2} \sim 400$
to 500~GeV, which is similar for all the studied values of $A_0$. In
this case, there is also a similar minimum of $\chi^2$ for the
focus-point region at $m_{1/2} \sim 200 \gev$. The increase in $\chi^2$
at small $m_{1/2}$ is due to $(g - 2)_\mu$ as well as $\Mh$, whereas the
increase at large $m_{1/2}$ is essentially due to $(g - 2)_\mu$. 
Contrary to the $\tb = 10$ case, $\Mh$ does not induce a large
difference for the various $A_0$ values, see the right plot in
\reffi{fig:Mh}. 
We note
that the overall minimum of $\chi^2 \sim 2$ is similar for both values
of $\tb$, and represents an excellent fit in each case.

\begin{figure}[tbh!]
\begin{center}
\includegraphics[width=.48\textwidth,height=4.5cm]{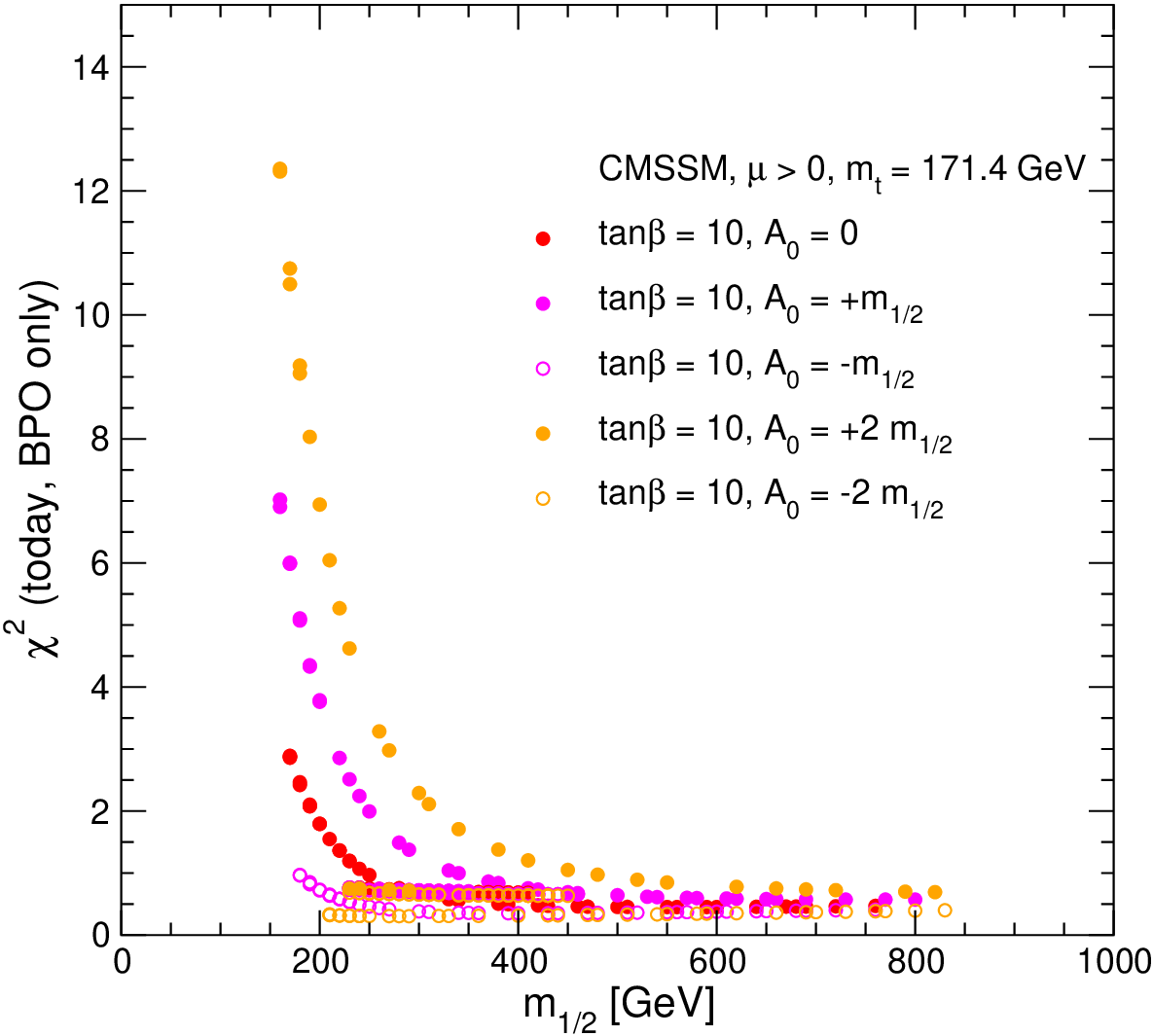}
\includegraphics[width=.48\textwidth,height=4.5cm]{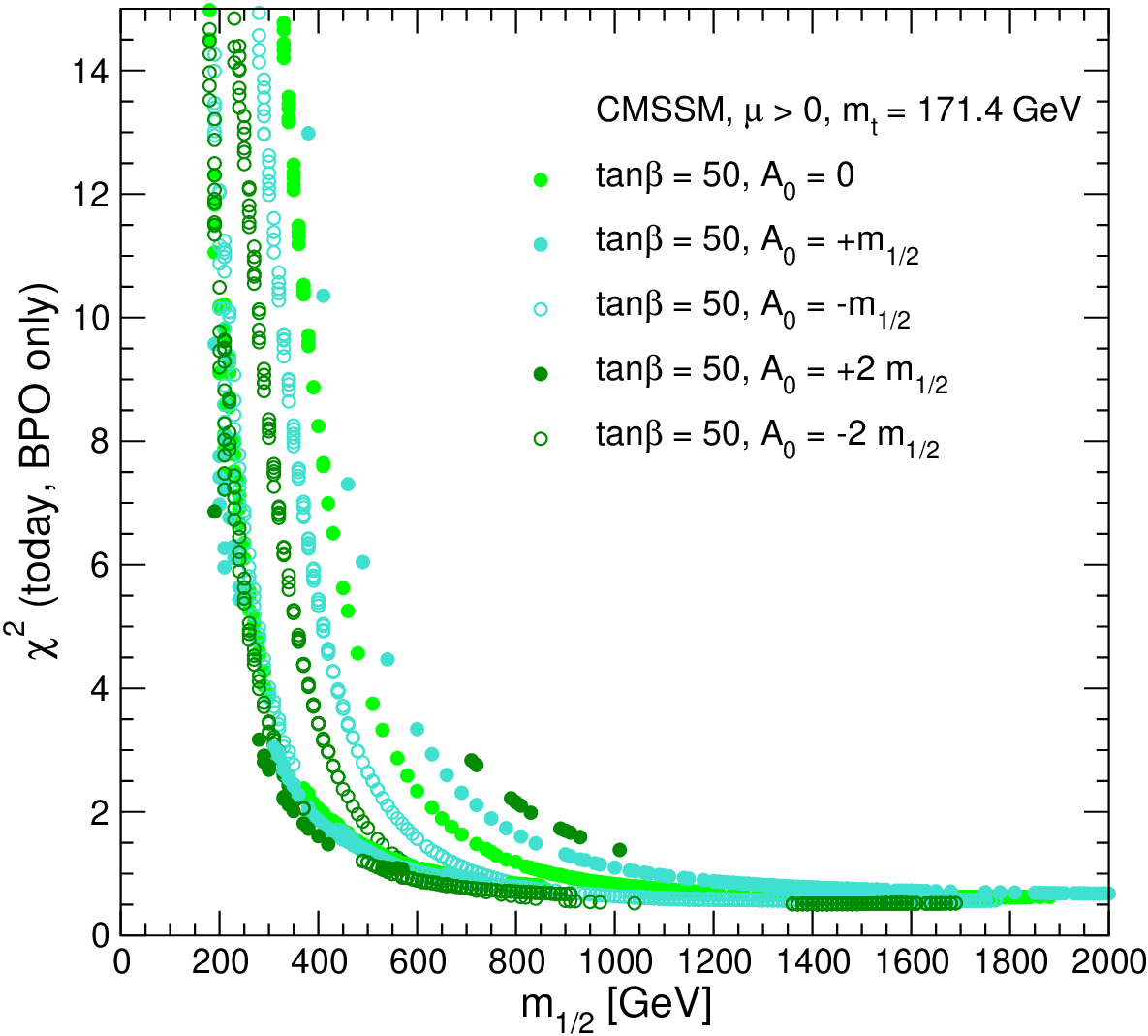}
\vspace{-0.5em}
\caption{%
The combined $\chi^2$~function for the $B$~physics observables
$\br(b \to s \ga)$, $\br(B_s \to \mu^+\mu^-)$, $\br(B_u \to \tau \nu_\tau)$
and $\De M_{B_s}$, evaluated in the CMSSM for $\tb = 10$ (left) and
$\tb = 50$ (right) for various discrete values of $A_0$.
}
\label{fig:chi_bpo}
\end{center}
\vspace{-2.0em}
\end{figure}

\reffi{fig:chi_bpo} shows the corresponding combined $\chi^2$ for the
BPO alone. For both values of $\tb$, these prefer large values of
$m_{1/2}$, reflecting the fact that there is no hint of any deviation
from the SM, and the overall quality of the fit is good. Very small values
of $m_{1/2}$ are disfavored, particularly in the coannihilation region
with $A_0 > 0$, mainly due to $b \to s \ga$. The focus-point region is
generally in very good 
agreement with the BPO data, except at very low $m_{1/2} \lsim 400 \gev$
for $\tb = 50$. 

\begin{figure}[bh!]
\begin{center}
\includegraphics[width=.48\textwidth,height=4.5cm]{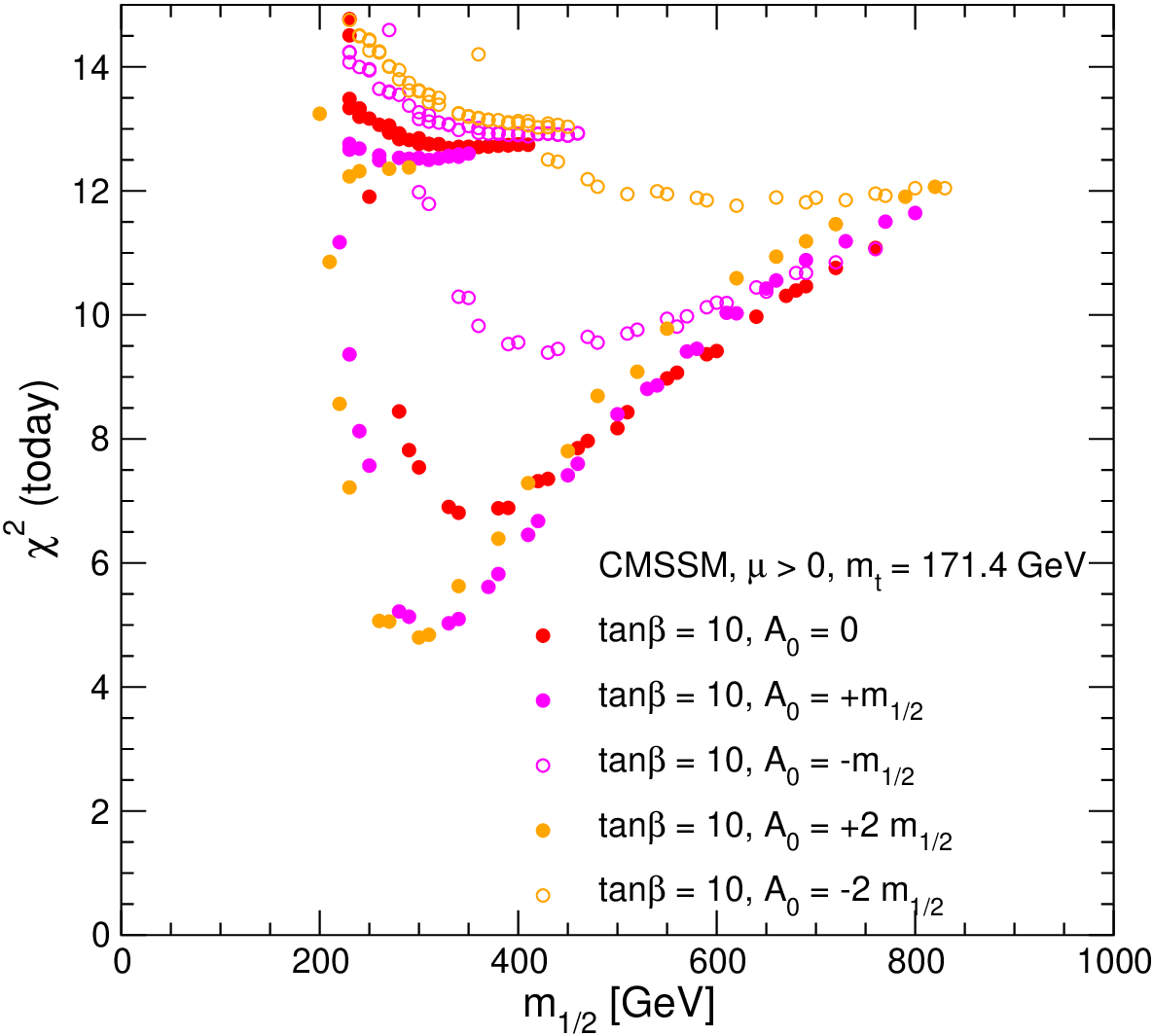}
\includegraphics[width=.48\textwidth,height=4.5cm]{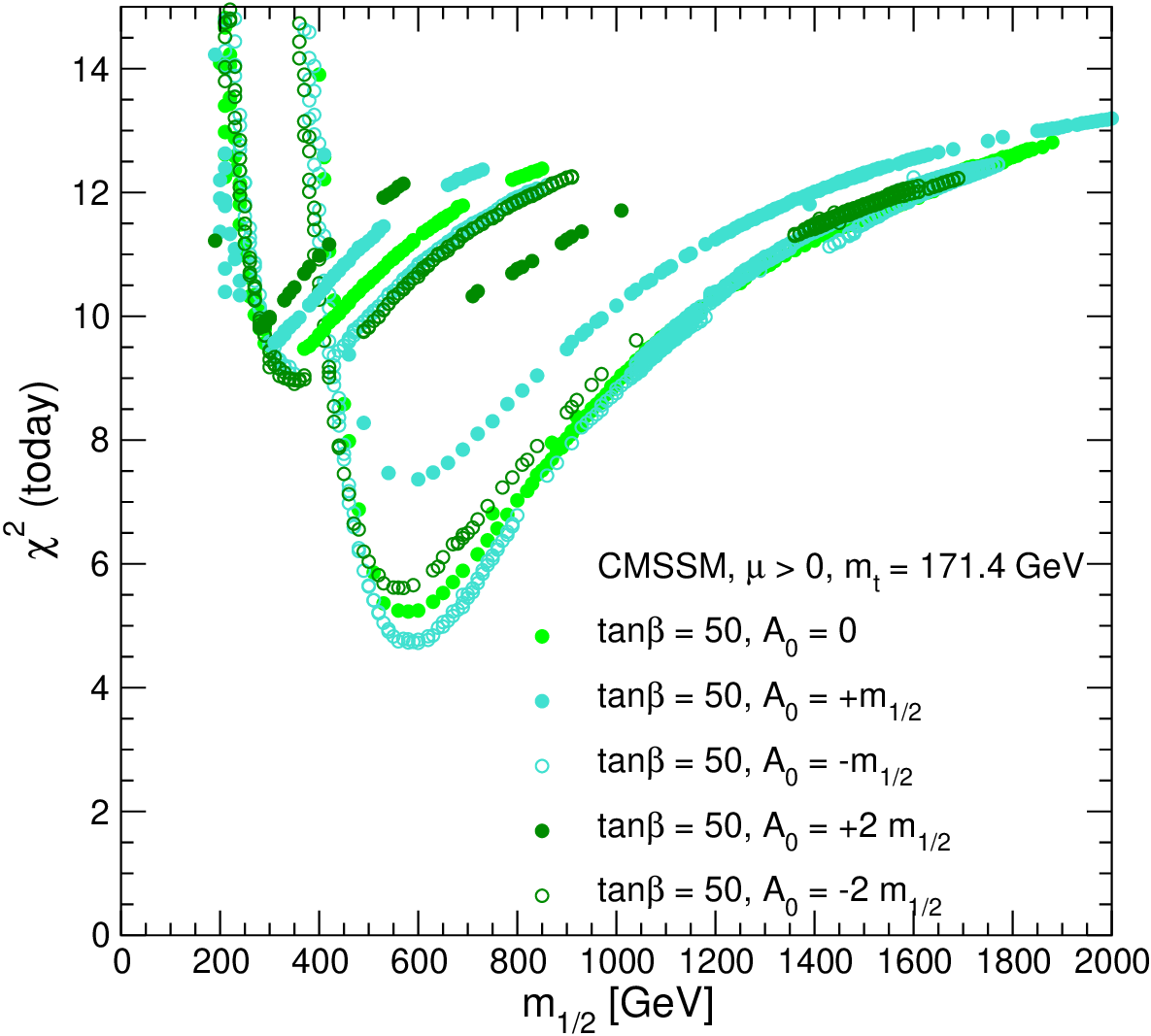}
\vspace{-0.5em}
\caption{%
The combined $\chi^2$~function for the EWPO and the BPO, 
evaluated in the CMSSM for $\tb = 10$ (left) and
$\tb = 50$ (right) for various discrete values of $A_0$. 
}
\label{fig:chi}
\end{center}
\vspace{-1em}
\end{figure}

\begin{figure}[htb!]
\begin{center}
\includegraphics[width=.48\textwidth,height=4.5cm]{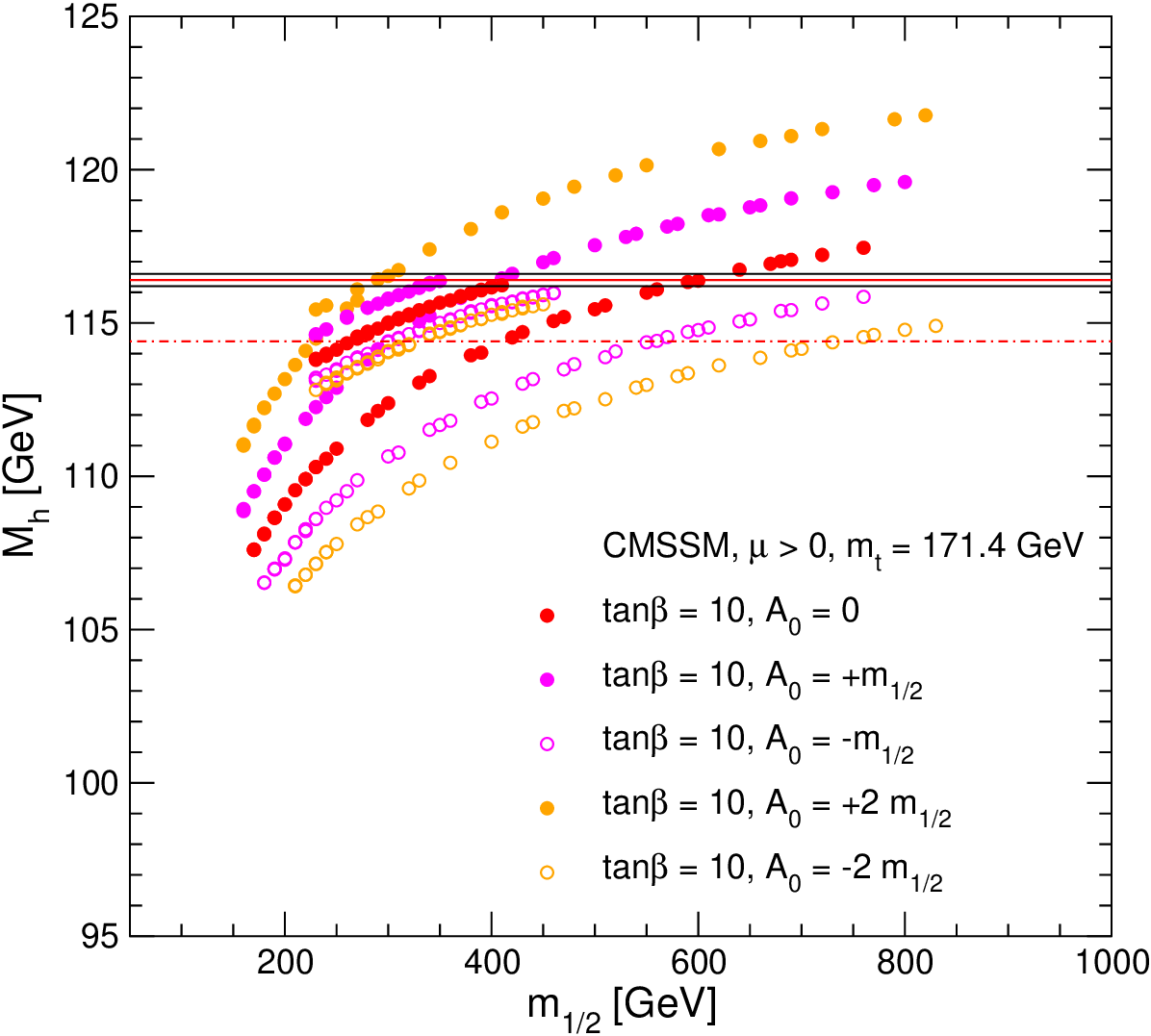}
\includegraphics[width=.48\textwidth,height=4.5cm]{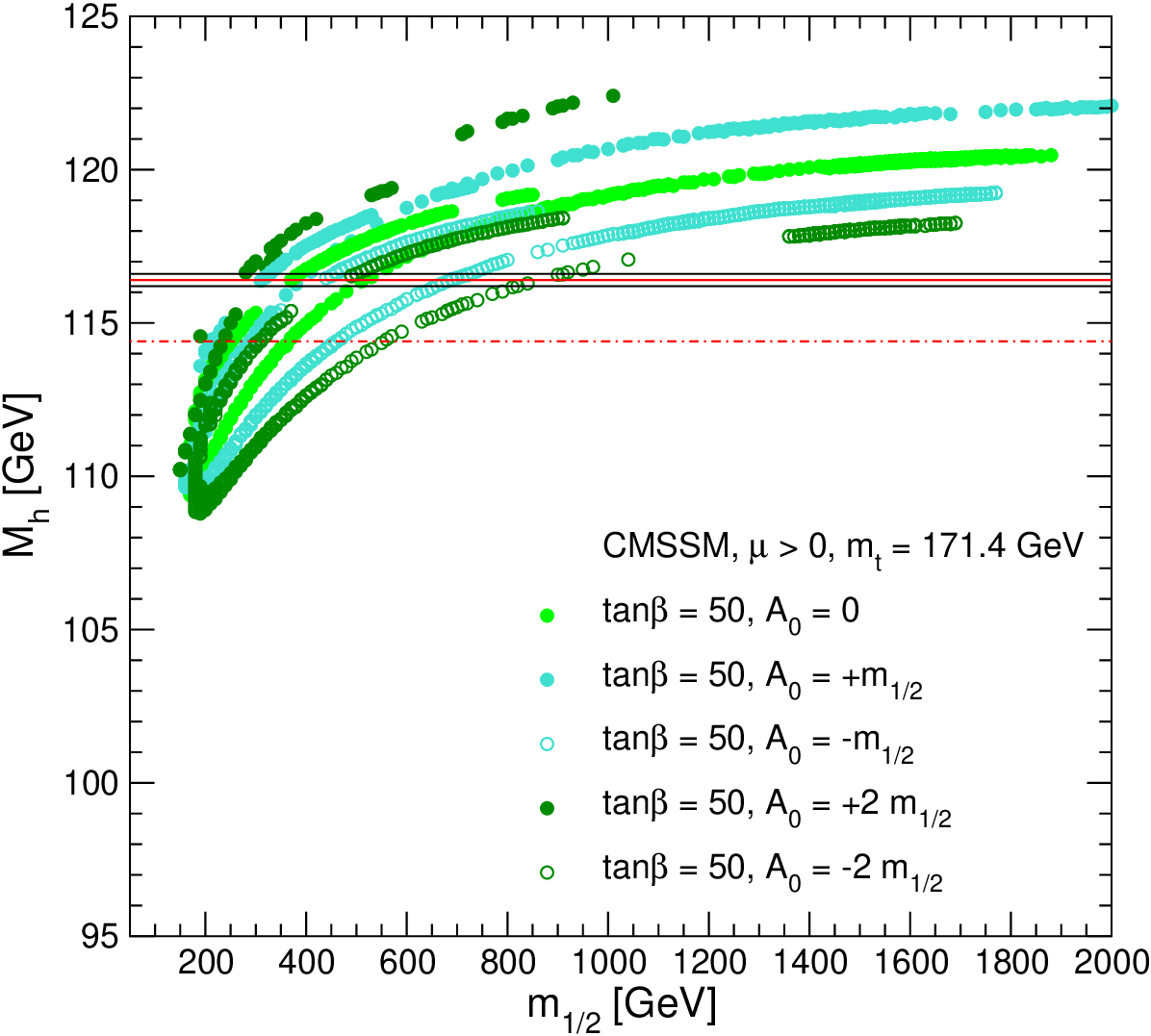}
\vspace{-0.5em}
\caption{
The CMSSM predictions for $\Mh$ as functions of $m_{1/2}$ with 
(a) $\tb = 10$ and (b) $\tb = 50$ for various $A_0$. 
We also show the present 95\%~C.L.\ exclusion limit of $114.4 \gev$ and
a hypothetical LHC measurement of $\Mh = 116.4 \pm 0.2 \gev$.
The results have been obtained with {\tt FeynHiggs}~\cite{feynhiggs}.
}
\label{fig:Mh}
\end{center}
\vspace{-1em}
\end{figure}

Finally, we show in \reffi{fig:chi} the combined $\chi^2$~values
for the EWPO and BPO, computed in accordance with \refeq{eq:chi2}.
We see that the global minimum of $\chi^2 \sim 4.5$
for both values of $\tb$. This is quite a good fit for the number of
experimental observables being fitted, and the $\chi^2/{\rm d.o.f.}$ is
similar to the one for the EWPO alone. This increase in the total
$\chi^2$ reflects the fact that the BPO
exhibit no tendency to reinforce the preference of the EWPO for small
$m_{1/2}$. However, due to the relatively
large experimental and theoretical errors for the BPO, no firm
conclusion in any direction can be drawn yet.
The focus-point region is disfavored for both values of $\tb$ by comparison
with the coannihilation region, though this effect is slightly less
important for $\tb = 50$. 
For $\tb = 10$, $m_{1/2} \sim 300 \gev$ and $A_0 > 0$ are preferred, whereas,
for $\tb = 50$, $m_{1/2} \sim 600 \gev$ and $A_0 < 0$ are preferred. This
change-over is largely due to the impact of the LEP $\Mh$ constraint for
$\tb = 10$ (see the left plot of \reffi{fig:Mh}) and the $b \to s \ga$
constraint for $\tb = 50$ (see Fig.~6 in \citere{ehoww}).
Corresponding mass predictions for the SUSY particles can be found in
\citere{ehoww}.



\section{Acknowledgments}

We thank J.~Ellis, K.A.~Olive, A.M.~Weber and G.~Weiglein for 
collaboration on the work presented here.
Work supported in part by the European Community's Marie-Curie Research
Training Network under contract MRTN-CT-2006-035505
`Tools and Precision Calculations for Physics Discoveries at Colliders'




\begin{footnotesize}


\end{footnotesize}


\begin{thebibliography}{99}
\bibitem{url} Slides: \\ 
\verb$ilcagenda.linearcollider.org/contributionDisplay.py?contribId=52&sessionId=69&confId=1296$

\bibitem{ehoww} J.~Ellis, S.~Heinemeyer, K.~Olive, A.M.~Weber and G.~Weiglein,
                {\em JHEP} {\bf 0708} (2007) 083
                [arXiv:0706.0652 [hep-ph]].

\bibitem{PomssmRep} S.~Heinemeyer, W.~Hollik and G.~Weiglein, 
                    {\em Phys.\ Rept.} {\bf 425} (2006) 265
                    [arXiv:hep-ph/0412214].

\bibitem{ehow3} J.~Ellis, S.~Heinemeyer, K.~Olive and G.~Weiglein,
                {\em JHEP} {\bf 0502} (2005) 013
                [arXiv:hep-ph/0411216].

\bibitem{ehow4} J.~Ellis, S.~Heinemeyer, K.~Olive and G.~Weiglein,
                {\em JHEP} {\bf 0605} (2006) 005 
                [arXiv:hep-ph/0602220].

\bibitem{other} J.~Ellis, K.~Olive, Y.~Santoso and V.~Spanos,
                {\em Phys.\ Rev.} {\bf D 69} (2004) 095004
                [arXiv:hep-ph/0310356];
                B.~Allanach and C.~Lester,
                {\em Phys.\ Rev.} {\bf D 73} (2006) 015013
                [arXiv:hep-ph/0507283];
                B.~Allanach,
                {\em Phys.\ Lett.} {\bf B 635} (2006) 123
                [arXiv:hep-ph/0601089];
                R.~de Austri, R.~Trotta and L.~Roszkowski,
                {\em JHEP} {\bf 0605} (2006) 002
                [arXiv:hep-ph/0602028];
                {\em JHEP} {\bf 0704} (2007) 084
                [arXiv:hep-ph/0611173];
                arXiv:0705.2012 [hep-ph];
                B.~Allanach, C.~Lester and A.~M.~Weber,
                {\em JHEP} {\bf 0612} (2006) 065
                [arXiv:hep-ph/0609295];
                B.~Allanach, K.~Cranmer, C.~Lester and A.~M.~Weber,
                arXiv:0705.0487 [hep-ph];
                O.~Buchmueller et al.,
                arXiv:0707.3447 [hep-ph].

\bibitem{LSPlargeTB} G.~Isidori, F.~Mescia, P.~Paradisi and D.~Temes,
                     {\em Phys.\ Rev.} {\bf D 75} (2007) 115019
                     [arXiv:hep-ph/0703035];
                     M.~Carena, A.~Menon and C.~Wagner,
                     {\em Phys.\ Rev.} {\bf D 76} (2007) 035004
                     [arXiv:0704.1143 [hep-ph]].

\bibitem{mt1714} E.~Brubaker et al. [Tevatron Electroweak Working Group],
                 arXiv:hep-ex/0608032, 
                 see:\\ {\tt tevewwg.fnal.gov/top/}~.

\bibitem{mt1709} Tevatron Electroweak Working Group,
                 arXiv:hep-ex/0703034.

\bibitem{focus} J.~Feng, K.~Matchev and T.~Moroi,
                {\em Phys. Rev. Lett.} {\bf 84} (2000) 2322
                [arXiv:hep-ph/9908309];
                {\em Phys. Rev.} {\bf D 61} (2000) 075005
                [arXiv:hep-ph/9909334];
                J.~Feng, K.~Matchev and F.~Wilczek,
                {\em Phys. Lett.} {\bf B 482} (2000) 388
                [arXiv:hep-ph/0004043];
                J.~Feng and K.~Matchev,
                {\em Phys. Rev.} {\bf D 63} (2001) 095003
                [arXiv:hep-ph/0011356].
                
\bibitem{WMAP} C.~Bennett et al.,
               {\em Astrophys. J. Suppl.} {\bf 148} (2003) 1
               [arXiv:astro-ph/0302207];
               D.~Spergel et al.\ [WMAP Collaboration],
               {\em Astrophys. J. Suppl.} {\bf 148} (2003) 175
               [arXiv:astro-ph/0302209];
               D.~Spergel et al.\ [WMAP Collaboration],
               arXiv:astro-ph/0603449.

\bibitem{LEPHiggs} LEP Higgs working group,
                   {\em Phys. Lett.} {\bf B 565} (2003) 61
                   [arXiv:hep-ex/0306033];
                   {\em Eur.\ Phys.\ J.} {\bf C 47} (2006) 547
                   [arXiv:hep-ex/0602042].

\bibitem{feynhiggs} S.~Heinemeyer, W.~Hollik and G.~Weiglein, 
                    {\em Comp. Phys. Commun.} {\bf 124} 2000 76
                    [arXiv:hep-ph/9812320];
                    {\em Eur. Phys. J.} {\bf C 9} (1999) 343
                    [arXiv:hep-ph/9812472];
                    G.~Degrassi, S.~Heinemeyer, W.~Hollik,
                    P.~Slavich and G.~Weiglein, 
                    {\em Eur. Phys. J.} {\bf C 28} (2003) 133
                    [arXiv:hep-ph/0212020];
                    M.~Frank, T.~Hahn, S.~Heinemeyer, W.~Hollik,  
                    H.~Rzehak and G.~Weiglein,
                    {\em JHEP} {\bf 0702} (2007) 047
                    [arXiv:hep-ph/0611326];
                    see: {\tt www.feynhiggs.de} .



\end{thebibliography}
\end{document}